\documentclass[letter,fleqn]{cas-dc}

\usepackage[numbers]{natbib}
\usepackage{amssymb}
\usepackage{subcaption}
\usepackage[utf8x]{inputenc}
\usepackage{color}
\usepackage{url}
\usepackage{graphicx} 

\def\tsc#1{\csdef{#1}{\textsc{\lowercase{#1}}\xspace}}
\tsc{WGM}
\tsc{QE}
\tsc{EP}
\tsc{PMS}
\tsc{BEC}
\tsc{DE}
\begin{document}
\let\WriteBookmarks\relax
\def\floatpagepagefraction{1}
\def\textpagefraction{.001}
\shorttitle{The S$\pi$RIT Time Projection Chamber}
\shortauthors{The S$\pi$RIT collaboration}
%\begin{frontmatter}

\title[mode = title]{The S$\pi$RIT Time Projection Chamber}

\author[addrs1,addrs2]{J. Barney}
\cormark[1]
\ead{barneyj@nscl.msu.edu}

\author[addrs1,addrs2]{J. Estee}

\author[addrs1,addrs2]{W.G. Lynch}
\cormark[2]
\ead{lynch@nscl.msu.edu}

\author[addrs3]{T. Isobe}

\author[addrs1]{G. Jhang}

\author[addrs3]{M. Kurata-Nishimura}

\author[addrs4]{A.B. McIntosh}

\author[addrs5]{T. Murakami}

\author[addrs1]{R. Shane}

\author[addrs1]{S. Tangwancharoen}

\author[addrs1,addrs2]{M.B. Tsang}

\author[addrs1]{G. Cerizza}

\author[addrs5,addrs3]{M. Kaneko}

\author[addrs7]{J.W. Lee}

\author[addrs1,addrs2]{C.Y. Tsang}

\author[addrs1]{R. Wang}

\author[addrs1]{C. Anderson}
\author[addrs3]{H. Baba}
\author[addrs8]{Z. Chajecki}
\author[addrs8]{M. Famiano}
\author[addrs1]{R. Hodges-Showalter}
\author[addrs7]{B. Hong}
\author[addrsTohoku]{T. Kobayashi}
\author[addrs9]{P. Lasko}
\author[addrs9]{J. {\L}ukasik}
\author[addrs3]{N. Nakatsuka}
\author[addrs4]{R. Olsen}
\author[addrs3]{H. Otsu}
\author[addrs9]{P. Paw{\l}owski}
\author[addrs10]{K. Pelczar}
\author[addrs11]{W. Powell}
\author[addrs3]{H. Sakurai}
\author[addrs1]{C. Santamaria}
\author[addrs1]{H. Setiawan}
\author[addrs3]{A. Taketani}
\author[addrs1]{J.R. Winkelbauer}
\author[addrs13]{Z. Xiao}
\author[addrs4]{S.J. Yennello}
\author[addrs1]{J. Yurkon}
\author[addrs13]{Y. Zhang}
\author{and the S$\pi$RIT collaboration}

\address[addrs1]{National Superconducting Cyclotron Laboratory, East Lansing, MI 48824, USA}
\address[addrs2]{Department of Physics and Astronomy, Michigan State University, East Lansing, MI 48824 USA}
\address[addrs3]{RIKEN Nishina Center, Hirosawa 2-1, Wako, Saitama 351-0198, Japan}
\address[addrs4]{Cyclotron Institute, Texas A$\&$M University, College Station, TX 77843, USA}
\address[addrs5]{Department of Physics, Kyoto University, Kita-shirakawa, Kyoto 606-8502, Japan}
\address[addrs7]{Department of Physics, Korea University, Seoul 02841, Republic of Korea}
\address[addrs8]{Department of Physics, Western Michigan University, Kalamazoo, Michigan 49008, USA}
\address[addrsTohoku]{Department of Physics, Tohoku University, Sendai 980-8578, Japan}
\address[addrs9]{Institute of Nuclear Physics PAN, ul. Radzikowskiego 152, 31-342 Krak\'{o}w, Poland}
\address[addrs10]{Gran Sasso National Laboratory - INFN, Via G. Acitelli 22, 67100 Assergi, L’Aquila AQ, Italy}
\address[addrs11]{Department of Physics, University of Liverpool, Liverpool, Merseyside, L69 7ZE, UK}
\address[addrs13]{Department of Physics, Tsinghua University, Beijing 100084, PR China}

\begin{abstract}
The SAMURAI Pion Reconstruction and Ion-Tracker Time Projection Chamber (S$\pi$RIT TPC) was designed to enable measurements of heavy ion collisions with the SAMURAI spectrometer at the RIKEN Radioactive Isotope Beam Factory and provide constraints on the Equation of State of neutron-rich nuclear matter. The S$\pi$RIT TPC has a 50.5 cm drift length and an 86.4 cm $\times$ 134.4 cm pad plane with 12,096 pads that are equipped with the Generic Electronics for TPCs readout electronics. The S$\pi$RIT TPC allows excellent reconstruction of particles and provides isotopic resolution for pions and other light charged particles across a wide range of energy losses and momenta. Details of the S$\pi$RIT TPC are presented, along with discussion of the TPC performance based on cosmic ray and experimental data.
\end{abstract}


\maketitle

\label{S:1}
Placing constraints on the density dependence of the symmetry energy term of the nuclear Equation of State (EoS) is an important research objective for both nuclear physics and astrophysics. The SAMURAI Pion-Reconstruction and Ion-Tracker (S$\pi$RIT) Time Projection Chamber (TPC) was designed and constructed to be used as the primary tracking detector in a series of experiments intended to constrain the symmetry energy at about twice saturation density~\cite{SHA15}. Central collisions of heavy ion beams with anticipated energy ranges of 200--500 MeV/u result in the copious production of charged particles. The TPC design allows determining Particle IDentification (PID) of charged pions and light charged particles up to Li by measuring their momentum distributions and energy losses. The S$\pi$RIT TPC was commissioned without magnetic field in 2015, providing critical tests of the TPC and trigger detectors~\cite{JHA16}. In 2016, the S$\pi$RIT TPC was installed inside the Superconducting Analyzer for MUlti-particles from RAdioIsotope beams (SAMURAI) spectrometer~\cite{KOB13} at the RIKEN Radioactive Isotope Beam Factory (RIBF)~\cite{YANO07}. Soon after, a campaign of experiments impinging neutron-rich Sn beams on isotopically enriched Sn targets was successfully carried out.
This paper details the design and construction of the S$\pi$RIT TPC, and discusses how the design was influenced by the experimental requirements. The paper is organized as follows: A brief overview of the TPC and operating principle, followed by further detailed description of individual components, is provided in Section 1. The performance of the S$\pi$RIT TPC is demonstrated using selected data from the experimental campaign in Section 2. 

\begin{figure}
  \centering
    \includegraphics[width=1\linewidth]{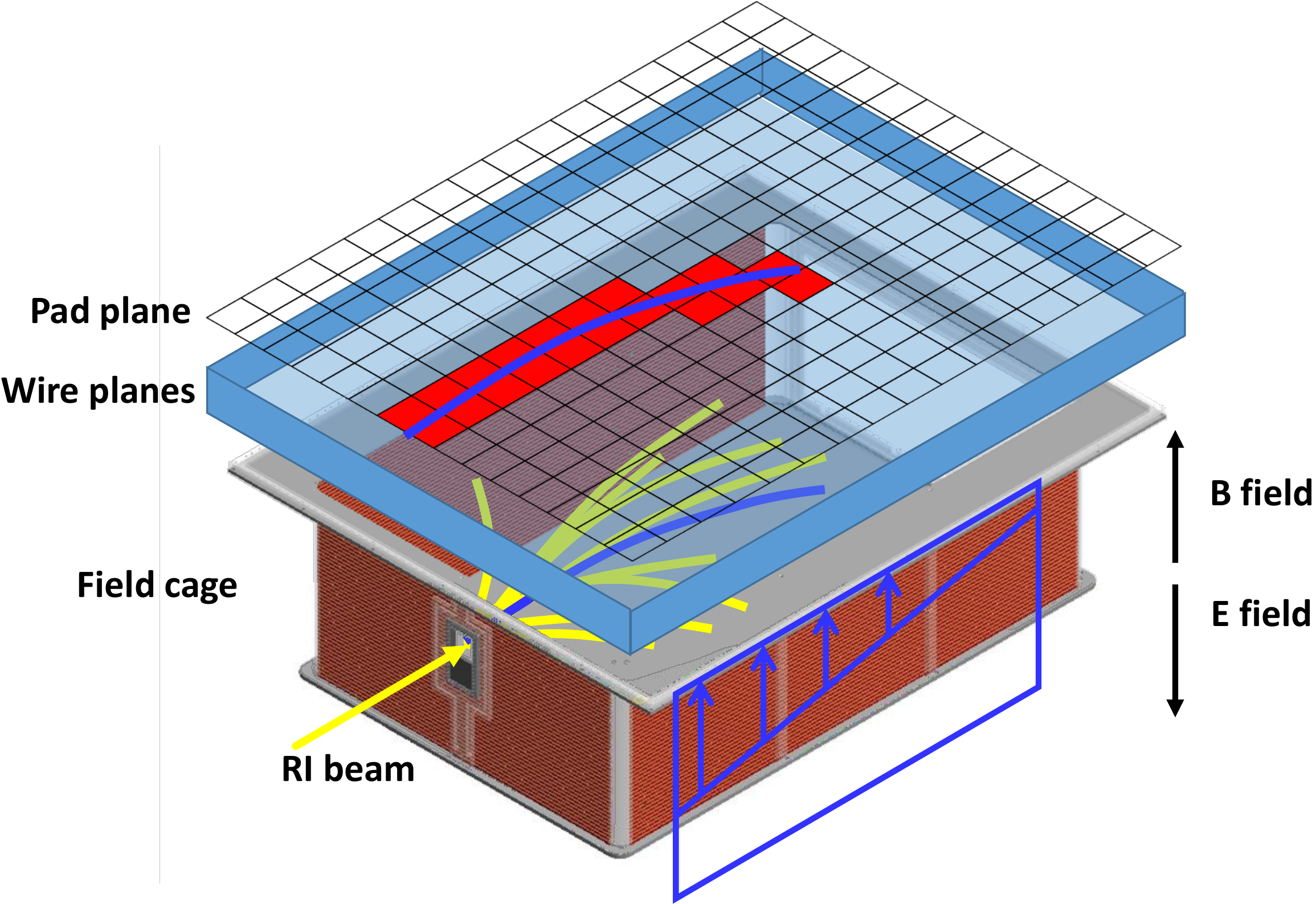}
\caption{Operation principle of the S$\pi$RIT TPC, modified from~\cite{SHA15}.}\label{OperationPrinciple}
\end{figure}

The design and operating principles of the S$\pi$RIT TPC are illustrated by Fig.~\ref{OperationPrinciple}. To avoid strong $\overrightarrow{E}\times\overrightarrow{B}$ drift velocity effects, the main electric and magnetic fields in a TPC are usually parallel, linking the geometry of a TPC design strongly to the geometry of its magnet. The S$\pi$RIT TPC is rectangular, designed to work with the SAMURAI dipole magnet~\cite{KOB13}. Its pad plane and wire planes lie perpendicular to the magnetic field and are located at the top of the detector. The field cage produces a uniform electric field anti-parallel to the magnetic field, and is filled with gas at just above atmospheric pressure to eliminate the possibility of air entering into the TPC. 

During the 2016 experimental campaign, the TPC was filled with P10 gas (90\% Argon, 10\% CH$_4$), and used to measure the collisions of Rare-Isotope (RI) beams with fixed targets located at the entrance of the field cage of the TPC. The focus of these experiments was to study central nucleus-nucleus collisions. The centrality of these collisions can be assessed by counting the number of emitted charged particles (multiplicity) emitted in these collision; this multiplicity generally increases as the impact parameter for the collision decreases. As these charged particles pass through the P10 gas within the field cage, they ionize molecules in the P10 gas mixture, creating electron-ion pairs. The positive ions drift downwards towards the cathode plate, while the electrons drift upwards towards the ground wire plane. The electrons undergo an avalanche process in the vicinity of the anode wire plane, which is located between the ground plane and pad plane. The high voltage potential between the ground and anode wire planes accelerates the drifting electrons, giving them sufficient energy to liberate additional electrons from the gas. The additional liberated electrons will also be accelerated, freeing more electrons. This sequence of ionization produces an avalanche of electrons, motivating the description of this region between the ground and anode wire planes as the avalanche region. The electrons terminate on the anode plane, and the motion of the positive ions produced in the avalanche region induces an image charge on the pad plane, creating a signal large enough to be amplified and digitized by the readout electronics.

\section{S$\pi$RIT Time Projection Chamber}

 Many aspects of the design of the S$\pi$RIT TPC resemble that of the EOS TPC~\cite{106592}, as both detectors measured heavy ion collisions in fixed target mode within a dipole magnet~\cite{1061019} of similar design. The S$\pi$RIT TPC operates within the SAMURAI dipole, which has a magnet diameter similar to that of the HISS dipole that housed the EOS TPC, but the SAMURAI dipole has a smaller usable gap of 75 cm. Table~\ref{tpc_properties} lists some of the S$\pi$RIT TPC operating parameters.

\begin{table}
\begin{center}
\caption{Properties of the S$\pi$RIT TPC\label{tpc_properties}}
 \begin{tabular}{ l l } 
\hline
Pad plane area & 134.4 cm $\times$ 86.4 cm\\
Number of pads & 12,096 (112 $\times$ 108) \\
Pad size & 1.2 cm $\times$ 0.8 cm \\
Drift distance & 50.5 cm \\
Electronics sampling frequency & 40 MHz \\
Signal shaping time & 117 ns \\
Electronic noise & 800 $q_e$ \\
Gas pressure & 1 atm \\
Typical gas composition & 90\% Ar + 10\% CH$_4$ \\
Gas gain & $\sim$1000\\
Electric field & 125 V/cm \\
Magnetic field & 0.5 T \\
Drift velocity & 5.5 cm/$\mu$s \\
Event rate & 50-60 Hz \\
Typical track multiplicity & 60 \\
\hline
\end{tabular}
\end{center}
\end{table}

\begin{figure}
\centering\includegraphics[width=1\linewidth]{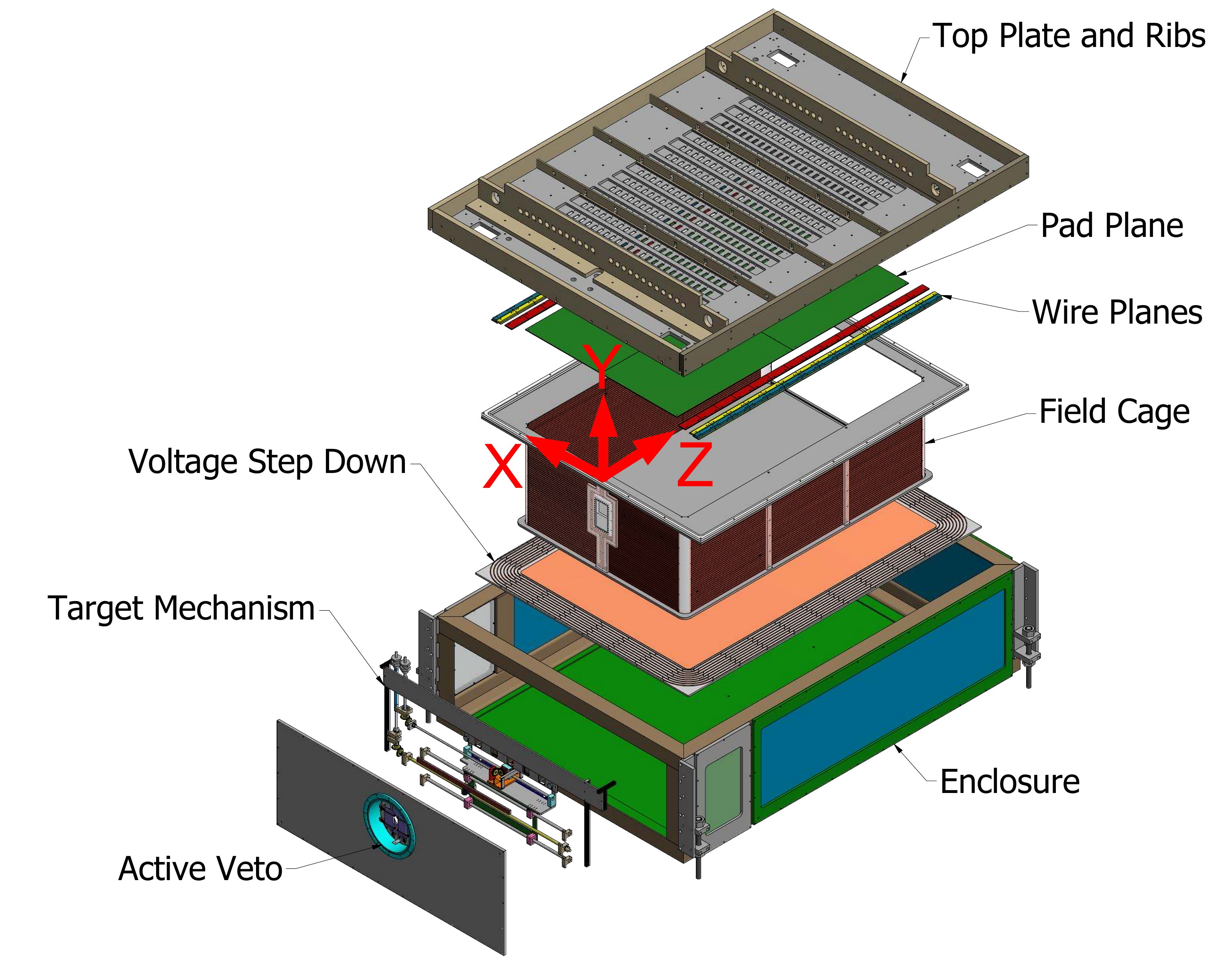}
\caption{Exploded view of the S$\pi$RIT TPC, with key components labeled.}
\label{TPCexplode}
\end{figure}

An exploded view of the S$\pi$RIT TPC is shown in Fig.~\ref{TPCexplode}, with key components labeled. The top plate provides the rigid mounting surface for the readout electronics, pad plane, wire planes, target mechanism, and field cage. The field cage is attached to the top plate. Unlike the EOS TPC, we position the wire planes at the top of the field cage, which helps to protect the pad and wire planes from falling objects. It also assists in the convective removal of the heat generated by the TPC readout electronics located on top of the TPC. The voltage step down, which safely interfaces the high voltage of the cathode to the ground of the enclosure, is located at the bottom of the enclosure. As discussed in more detail later, the voltage step down defines the electric field between the high voltage of the cathode with the outer enclosure, and is designed to safely allow much higher electric fields up to 400 V/cm. This makes it possible to use gases with lower drift velocities such as H$_{2}$ or He-CO$_{2}$ within the TPC. The first experimental campaign ran in fixed target mode, with metallic Sn target foils located 9 mm in front of the entrance window of the field cage, using P10 as the counter gas.

To facilitate the description of the TPC, we define a coordinate system of the TPC relative to the pad plane, with the origin and axis shown in Fig.~\ref{TPCexplode}. The $y$ dimension lies along the direction of the magnetic field, which points upward. The $z$ dimension lies along the beam axis, and the $x$ dimension is defined to be consistent with a right-handed Cartesian coordinate system.

\subsection{The Rigid Top Plate}
The rigid top plate is made of aluminum, with a set of aluminum ribs affixed to the sides. These ribs contribute to the rigidity of the top plate, and provide support for the weight of the ASIC and ADC (AsAd) front end electronic modules of the Generic Electronics for TPC (GET)~\cite{POLLACCO201881} system. The rigidity of the top plate ensures that the pad plane remains flat, and the distance from wire planes to pad plane remains fixed. The GET electronics provides 12 bit readout (4096 channels) of the pad signals. This is important for achieving the wide dynamic range in energy loss required for the scientific program of the S$\pi$RIT TPC. Both complete readout of all channels or partial readout of individual thresholds above their individual channels can be performed. Due to the high charged particle multiplicities of interesting events in the first experimental campaign, however, there was no advantage to the partial readout mode and all channels were read for each event instead. A detailed description of the implementation of the GET electronics with the S$\pi$RIT TPC is provided in Ref.~\cite{ISOBE201843}.

\subsection{Pad Plane and Wire Planes}
The S$\pi$RIT TPC utilizes a Multi-Wire Proportional Chamber (MWPC) readout, with an anode and ground plane defining an avalanche region which produces signals on a charge-sensitive pad plane. An additional wire plane, referred to as the gating grid, is used to control the flow of electrons into the avalanche region as well as the back-flow of positive ions into the larger field cage volume. The pad plane and wire planes are mounted on the bottom of the rigid top plate, which corresponds to the top of the field cage, as shown in Fig.~\ref{TPCexplode}. 

\begin{table*}
\begin{center}
\caption{Wire plane properties.\label{wire_plane_table}}
 \begin{tabular}{c c c c c c c } 
Plane & Wire & Diameter & Pitch & Distance to & Tension & Total\\
 & material & (\textmu m) & (mm) & pad plane (mm) & (N)  & wires\\
 \hline
Anode & Au-plated W & 20 & 4 & 4 & 0.5 & 364 \\
Ground & BeCu & 75 & 1 & 8 & 1.2  & 1456 \\
Gating & BeCu & 75 & 1 & 14 & 1.2  & 1456 \\
 \hline
\end{tabular}
\end{center}
\end{table*}

This choice of multiwire gas amplification plane versus a GEM~\cite{sauli1997gem} or MICROMEGAS~\cite{giom1996mmegas} gas amplifier has both advantages and disadvantages. One disadvantage is that wire planes require considerable labor to construct. Another is the requirement for a gating grid to prevent backstreaming of positive ions produced at the anode wires during the gas amplification process. Opening and closing this gating grid introduces deadtime. Neither disadvantage has proven problematic for the S$\pi$RIT TPC experimental program. On the other hand, there are advantages to our multi-wire gas amplification choice because its wire amplification geometry induces useful image charges over both the central pad and the adjacent pads. Using the pad distribution function and the charge deposited on adjacent pads one can determine track centroids with mm resolution even with pads that have lateral dimensions of the order of 1 cm. Furthermore, since the induced charges on the neighboring pads in the S$\pi$RIT TPC are roughly a factor of five smaller than the charge induced in the pads directly below the track, the pad response function can be used to extend the energy loss measurement dynamic range by a factor of five~\cite{ESTEE2019162509}. With the software algorithm described in Ref.~\cite{ESTEE2019162509}, which uses the experimental pad response function and the 4096 channel resolution of the GET electronics (discussed below), the S$\pi$RIT TPC is able to achieve the dynamical range requirements for measurements of central collisions of heavy ion beams with anticipated energy ranges of 200--500 MeV/u~\cite{ESTEE2019162509}.

\begin{figure}
\centering\includegraphics[width=1\linewidth]{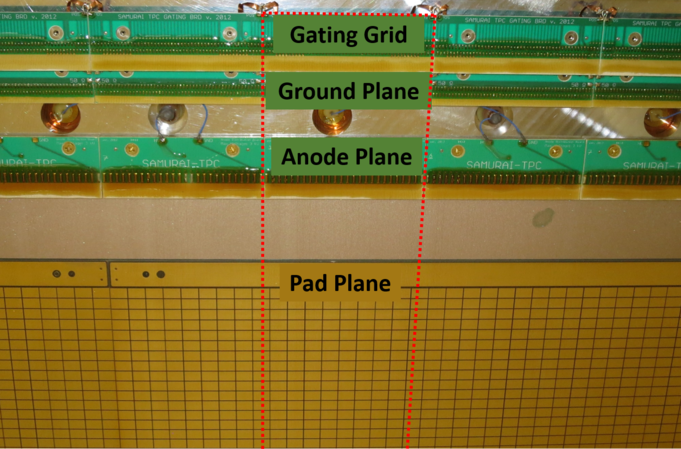}
\caption{Photograph of the pad plane and wire planes. The red dashed line shows the outline of a single wire plane section.}
\label{ppWP}
\end{figure}

\begin{figure}
\centering\includegraphics[width=1\linewidth]{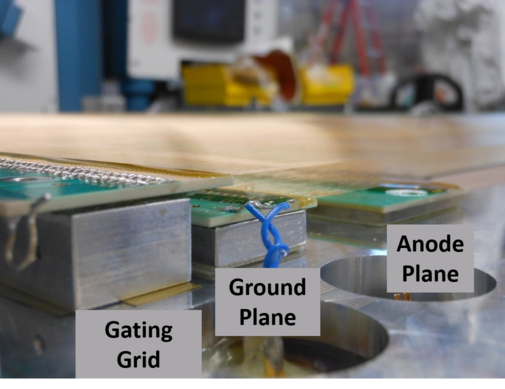}
\caption{Photograph of the wire planes, from the side. Each plane is labeled.}
\label{WPside}
\end{figure}

The physical properties of each wire plane are listed in Table~\ref{wire_plane_table}. The anode plane, made with 20 \textmu m diameter gold plated tungsten wires~\cite{LumaMetall}, is 4 mm below the pad plane, with a pitch of 4 mm between wires. The ground plane, made with 75 \textmu m diameter beryllium copper wires~\cite{CaliFine}, is 8 mm below the pad plane, with a pitch of 1 mm between wires. The gating grid plane, made with 75 \textmu m diameter beryllium copper wires~\cite{CaliFine}, is 14 mm below the pad plane with a pitch of 1 mm between wires. The wire planes are separated into 14 sections, with each section spanning 10.4 cm along the beam axis, and stretched across the width of the pad plane. Each section contains 26 anode wires, 104 ground wires, and 104 gating grid wires. The wires are positioned to be symmetric across the center of each pad, with every third anode wire centered below a pad, and every fourth ground and gating grid wire centered below an anode wire. Fig.~\ref{ppWP} shows the outline of a single wire plane section with a dashed line, and Fig.~\ref{WPside} shows the wire planes from the side. 

The ground plane is normally grounded to the TPC enclosure. We implement a switch to remove this ground connection in order to pulse the ground plane. Pulsing the ground plane induces signals uniformly in the pad plane, allowing the electronic gain and the zero offset of the pad plane electronics to be determined.

The amplification of charge, or gain, is described by the first Townsend coefficient $\alpha$, which depends on the gas properties and electric field within the amplification region~\cite{blum2008particle}. This is defined by the voltage on the anode wires, the distance between anode and pad plane, and the distance between anode plane and ground plane. The dependence of $\alpha$ on gas choice and density can be determined using simulation software such as GARFIELD~\cite{veenhof1998garfield}, or by comparison to measurements with similar electric fields. For the S$\pi$RIT TPC with P10 gas at atmospheric pressure and anode wires at 1460 V, the gas gain was calculated to be about 1000 using GARFIELD. The observed signals in the pad plane are compatible with these calculations. The anode wire planes are biased using two iseg EHS 8630p-F modules, mounted in an ECH 238 chassis. With these modules, we can also monitor the current in the anode wires.

The gating grid is used to prevent the amplification of charge from uninteresting events which can induce background, accelerate detector aging, and add to space charge effects. The amplification of electrons produces a large number of positive ions that move away from the anode plane towards the drift region. These ions move much slower than the electrons, and can accumulate within the drift region, distorting the electric field and affecting the track reconstruction accuracy. The back flow of these positive ions can be mitigated if they are captured on wires shortly after leaving the avalanche region. Detector aging occurs when the amplification of electrons produces negative polymers from impurities in the gas. Such polymers will collect on the anode wires, increasing their effective radius, and reducing the gas gain and deteriorating the performance of the TPC~\cite{BLINOV200395}. By preventing electron amplification for events which do not satisfy the physics trigger, the production of space charge and the detector aging effects are reduced. This is accomplished by the gating grid wire plane. 

The capacitance of the gating grid plus biasing cables is approximately 26.5 nF. If bias is applied through a driver supply and transmission line of 50 Ohm impedance, the RC time constant for discharging the gating grid and opening would be approximately 1.3 $\mu $s. This opening time would be significant compared to the total drift time~9.5 $\mu $s in the field cage of the TPC. Such a long opening time would cause electrons from the tracks in much of the upper part of the field cage to be deposited on the gating grid before it opens; that portion of the tracks would not be recorded as data. To address this problem, we designed a driver which charged and discharged the gating grid through a transmission lines with impedance of Z<2 ohm. With this, we opened the S$\pi$RIT TPC gating grid in approximately 350 ns. The opening and closing of the gating grid is controlled with a bipolar gating grid driver, developed specifically for the S$\pi$RIT TPC~\cite{TAN17}. The gating grid is biased by two Kikusui PMC500-0.1A power supplies, and the power to the gating grid driver is supplied using a Kikusui PMC18-3A power supply.

\begin{figure}
\begin{subfigure}[b]{1\linewidth}
%\vspace*{10pt}
 \includegraphics[width=1\linewidth]{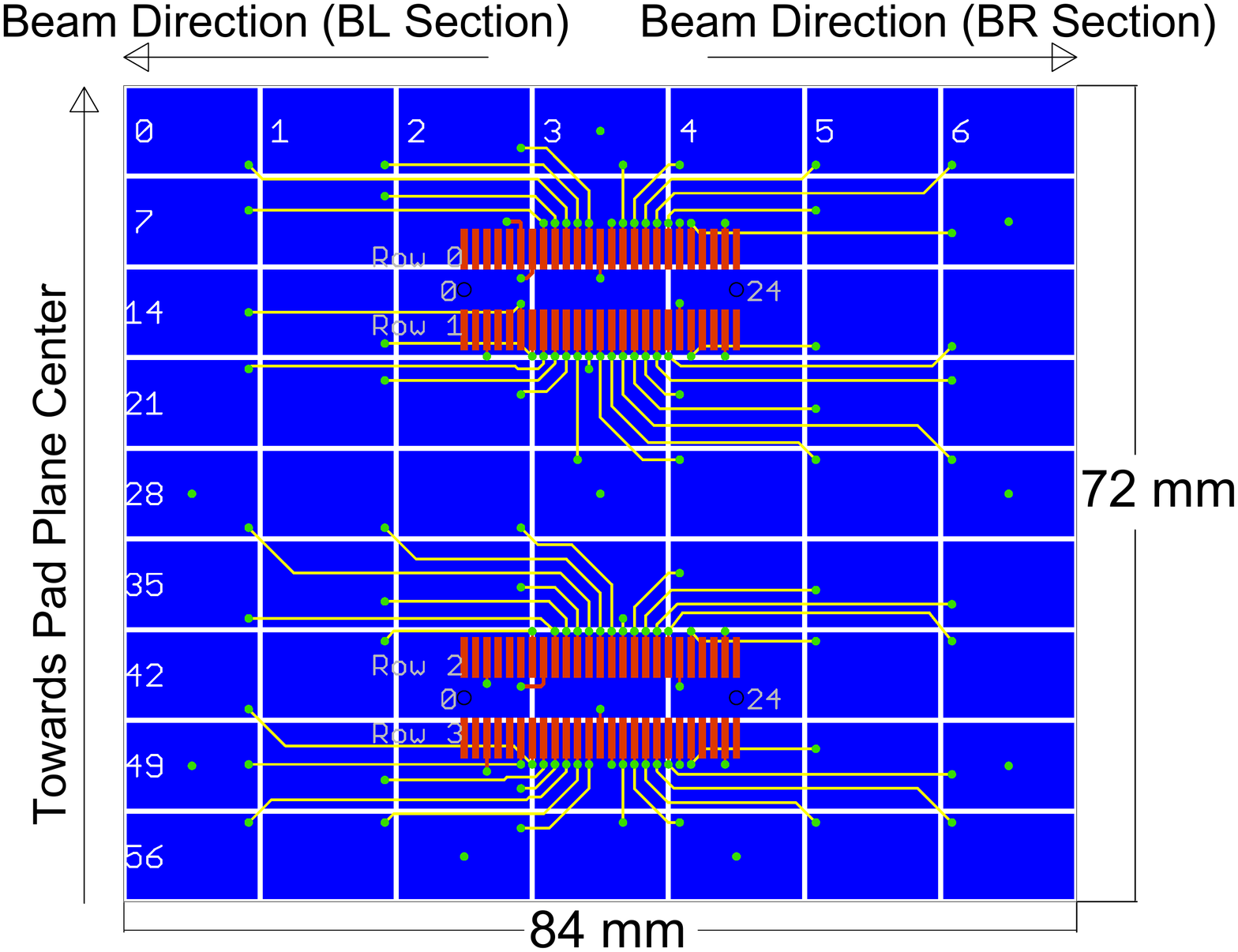}
 \caption{}
 \label{ppUnit}
\end{subfigure}
\begin{subfigure}[b]{1\linewidth}
 \includegraphics[width=1\linewidth]{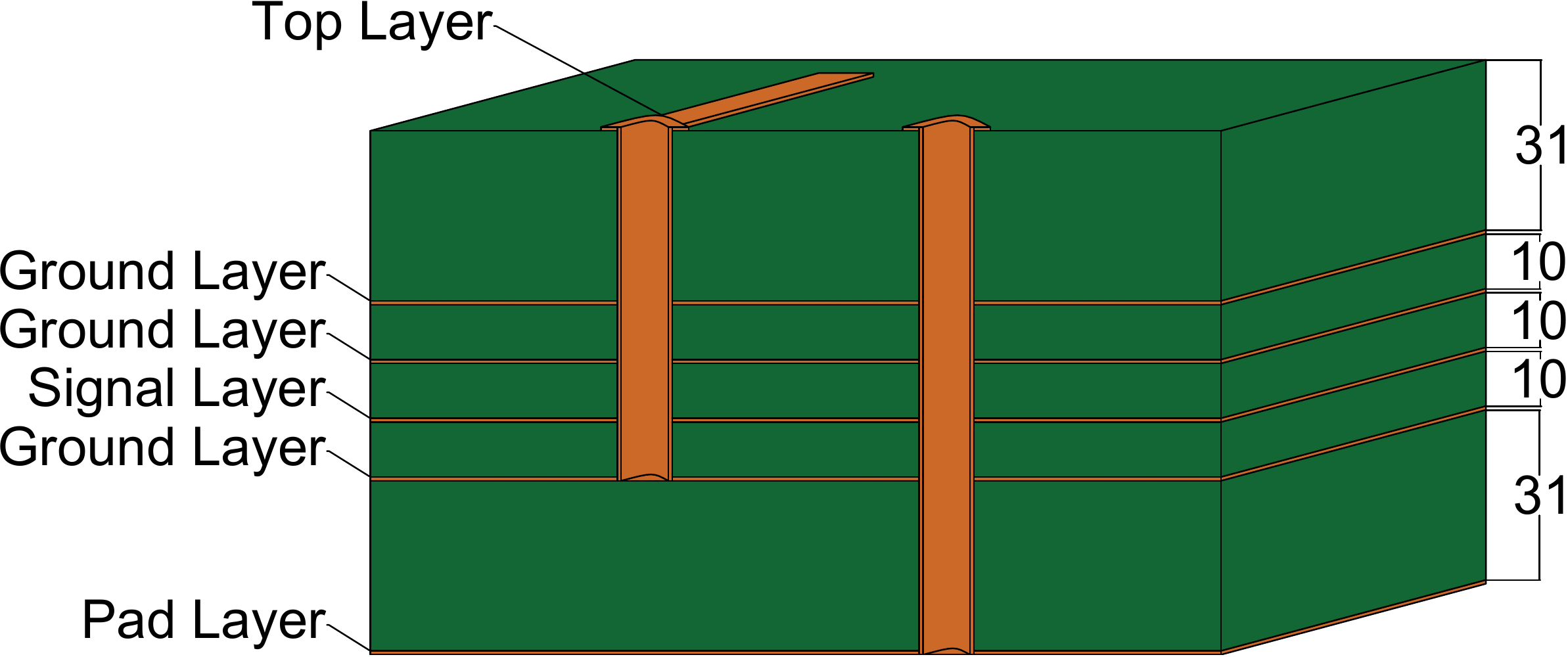}
 \caption{}
 \label{ppCS}
\end{subfigure}
\caption{The pad plane ``unit cell" (a) and the pad plane layer cross section (b). Layer thickness indicated in mil (1 mil = 0.001").}\label{ppDetails}
\end{figure}

The pad plane defines the readout area of the TPC, and is epoxied onto the bottom surface of the top plate. It contains 12,096 gold plated pads, spanning 1.2 cm $\times$ 112 pads = 134.4 cm in the beam direction, and spanning 0.8 cm $\times$ 108 pads = 86.4 cm in the direction perpendicular to both the beam and magnetic field. The pads are arranged in ``unit cells" (shown in Fig.~\ref{ppUnit}) consisting of 63 pads each. The charge sensitive portion of each pad is 11.5 mm long and 7.5 mm wide, with a 0.5 mm gap between pads. To avoid the possibility of problems with damaged connectors on the pad plane, each unit cell is read out through two Samtec FSI-125-10-L-D-AD push-type connectors. This avoids mounting connectors with easily damaged movable parts on the circuit boards; it only requires the passive contacts of the stationary board connections that are shown in red in Fig.~\ref{ppUnit}. The push-type connectors are mounted on rigid Printed Circuit Boards (PCBs), which connect to a protective circuit board (known as the ZAP board) through short section of ribbon cable. The ZAP board provides both protection from large signals (such as sparks) which could damage the readout electronics, and adapts signals to the high density connectors on the Amplifier, Shaper and Amplitude to Digital converter (ASAD) board of the GET readout electronics. To reduce noise and crosstalk issues, the pad plane was constructed from four 6-layer PCBs, with the cross section shown in Fig.~\ref{ppCS}. The typical capacitance between pads and the ground was measured to be 15 pF. Including the capacitance of the ZAP board and cable connections, the total capacitance presented to the ASAD preamp on each pad is approximately 25-30 pF. The signals from the pads were channeled to the Samtec connectors through 384 openings in the top plate. Each Samtec connector was inserted through the hole to mate with matching contacts on the back side of the pad plane circuit boards.

\begin{figure}
\centering\includegraphics[width=1\linewidth]{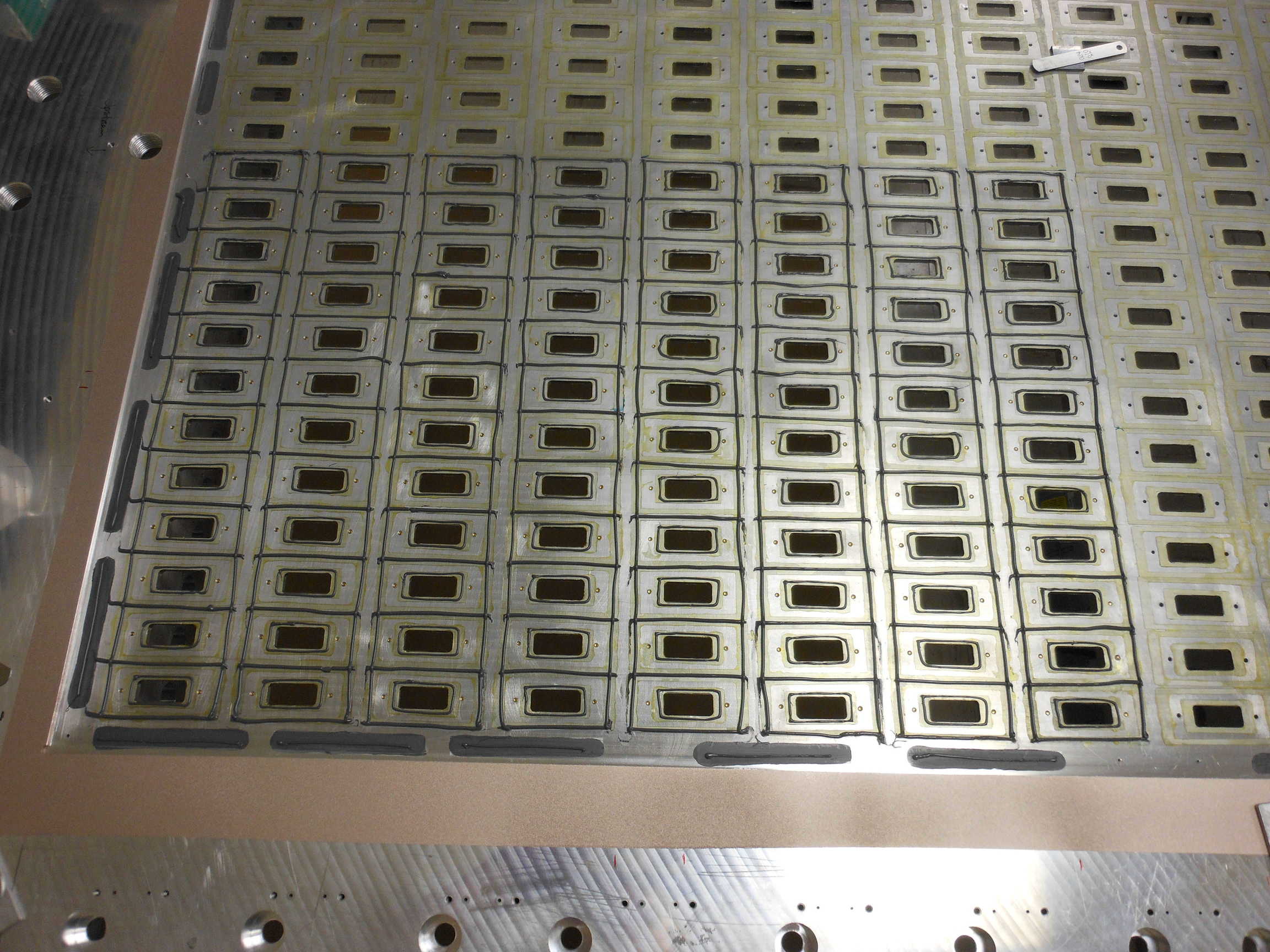}
\caption{Photograph of the top plate and glue pattern prior to gluing the pad plane to the top plate. The dark rectangular lines are the Araldite 2013 glue beads.}
\label{ppGluing}
\end{figure}

Due to the overall size and complexity, the pad plane was fabricated with four halogen-free G10 PCBs. Each PCB is 69.2$\times$45.2 cm, including a small isolation area around the perimeter of the pad plane. To bond the PCBs to the pad plane while achieving a pressure tight seal, lines of Araldite 2013 epoxy were applied in two concentric rectangular patterns around each of the 384 openings as shown in Fig.~\ref{ppGluing}. Then the PCBs were lowered onto precision shims and held in position against the rigid top plate using a precision-machined vacuum table, ensuring the flatness of the pad plane. After curing, the pad plane was measured with a laser alignment system, indicating uniform flatness within 125 \textmu m. During tests with P10 gas, some leaks were detected with a combustible gas detector. Such leaks were sealed by injecting EZ-poxy through pre-existing screw holes (needed for mounting the Samtec connectors) into the space between the concentric rectangular Araldite 2013 patterns for each of the openings through which the gas was observed to be leaking. Further details of the pad plane assembly procedure can be found in Ref.~\cite{barney2019dissertation}.

\subsection{Field cage}

\begin{figure}
\centering\includegraphics[width=1\linewidth]{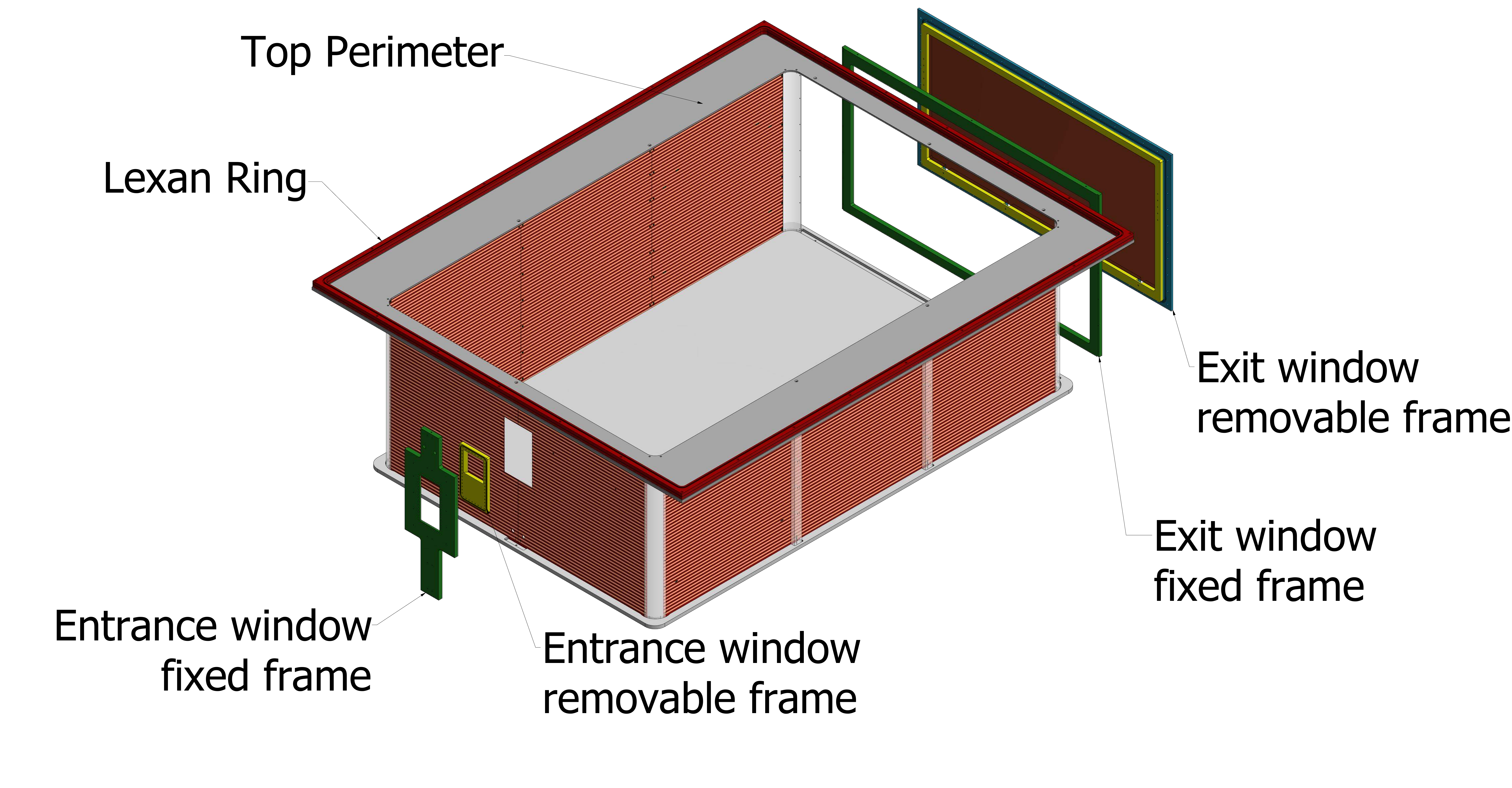}
\caption{Schematic view of the field cage. The entrance and exit windows are shown offset from the walls. }
\label{fc_window_and_walls}
\end{figure}

The field cage is the heart of the TPC, defining the detection volume in which particle tracks can be reconstructed. The field cage has interior dimensions of 145 cm in length, 97 cm in width, and the distance  between cathode and pad plane is 51.3 cm. The field cage is shown in Fig.~\ref{fc_window_and_walls}, with the windows and frames offset from the field cage walls. The entrance and exit windows are much thinner than the field cage walls, allowing particles to enter and exit the field cage with minimal energy loss. Equipotential strips on the walls and windows define a uniform electric field between the cathode plate, the top perimeter and the gating grid. The field cage is sealed by an O-ring surface at the interface between field cage and top plate, making it a gas tight detection volume. It is separated from the insulation gas volume between the field cage and the gas-tight outer enclosure; allowing the possibility of different drift and insulation gas compositions. Separating drift and insulation volumes allows the S$\pi$RIT TPC to  be run as an "active target" with a pure target gas such as H$_{2}$ within the field cage inducing the reaction, and an insulation gas such as N$_{2}$ with higher dielectric strength filling the insulation volume where higher electric fields are present.

The upstream and two side walls of the field cage are made with 1.6 mm thick 2-layer G-10 PCBs, which have a precise uniform pattern of copper strips on the inside and outside, forming the equipotential strips that define the electric field. Each side wall is made with 3 PCBs, which are aligned and joined together using polycarbonate bars. The front wall is made with 2 PCBs, with a polycarbonate window frame in the middle. The front entrance window is 5.73 cm wide by 7 cm tall, and made of 4 \textmu m thick poly p-phenylene terephthalamide (PPTA). The rear wall is made entirely of an exit window on a polycarbonate frame. The exit window, made of 125 \textmu m thick polyamide, is 80.8 cm wide by 38.9 cm tall and forms the downstream wall of the field cage. Aluminum strips are evaporated onto the windows to align with the copper strips on the PCBs. Both the upstream and downstream windows are removable to facilitate repair or replacement. The four corners are formed with rounded G-10, eliminating sharp corners in the geometry, which can cause spark formation. The window frames and rounded corners have equipotential strips made with conductive paint. For this and other equipotential surfaces formed with conductive paint, we used Parker Cho-shield 599, which is a two component, copper filled, conductive epoxy paint formulated to provide electromagnetic interference shielding and electrical grounding on insulating surfaces. Twenty-four quartz windows are installed in the field cage to allow a laser calibration of the drift velocity within the field cage, although these laser windows have not yet been employed for that purpose. 

\begin{figure}
\centering\includegraphics[width=1\linewidth]{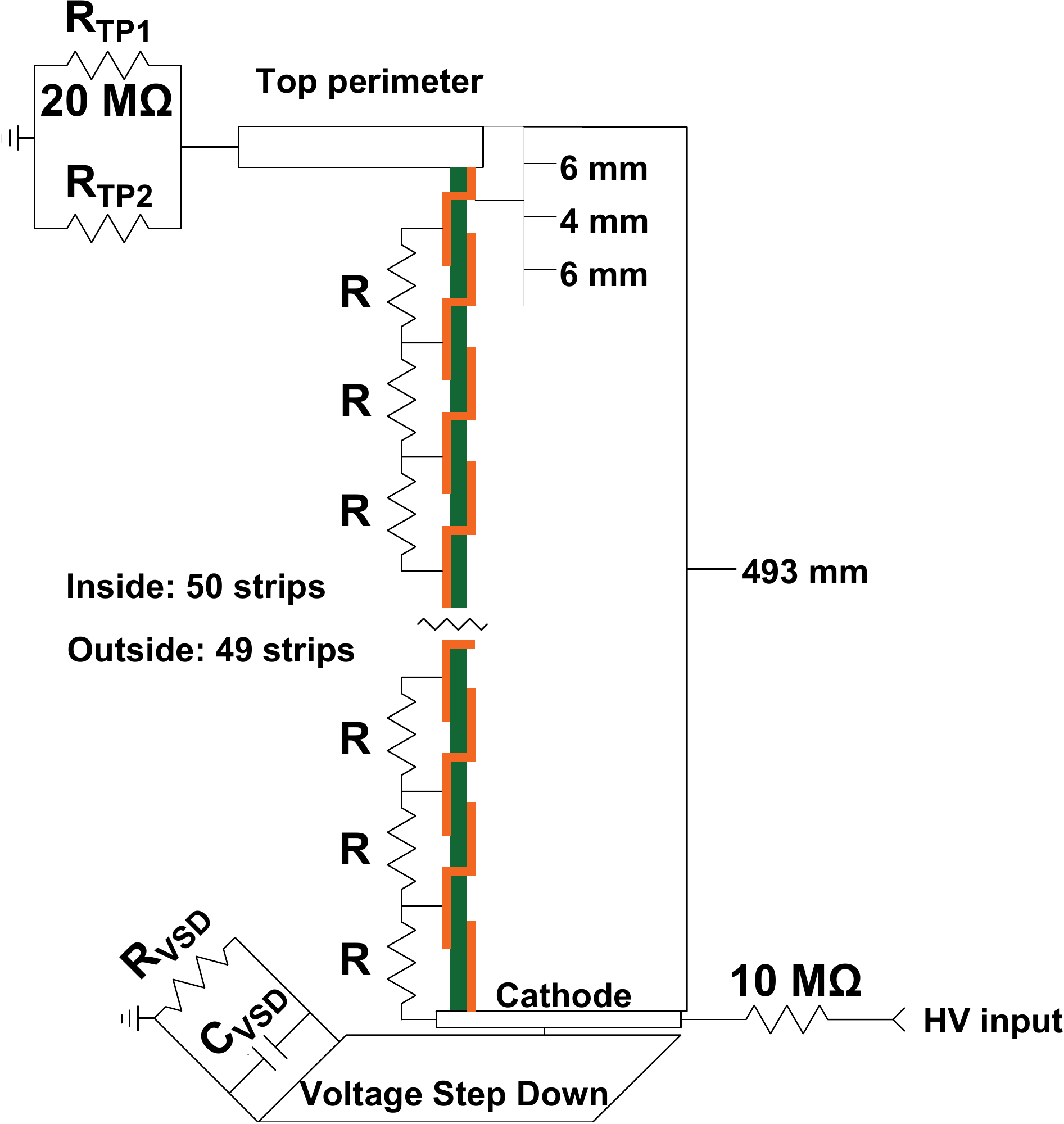}
\caption{Schematic view of field cage circuit layout.}
\label{fc_strips}
\end{figure}

The 6 mm equipotential strips on the field cage are spaced from the cathode to the top perimeter with 1 cm pitch, with sequential strips connected across the 4 mm gap by resistors. There are 50 strips on the inside of the field cage, and 49 on the outside. This produces a uniform electric field that is directed downwards towards the cathode. The layout is shown in Fig.~\ref{fc_strips}, with 49 resistors with resistance R=5 M$\Omega$, as well as R$_{\textrm{VSD}}$=700 M$\Omega$, and R$_{\textrm{TP2}}$=19.77 M$\Omega$. The overlapping structure of strips was chosen to effectively shield the pad plane from external noise. The capacitance C$_{\textrm{VSD}}$=4 nF in the circuit is dictated by the thickness and area of the voltage step down and is therefore an inherent property of the voltage stepdown. Combined with the resistor chain in Fig.~\ref{fc_strips}, the circuit effectively filters out oscillatory noise from the HV supply. High voltage is applied to the cathode, and the strength of the electric field is adjusted by varying the cathode voltage. For the S$\pi$RIT experiments of 2016, the cathode voltage was set at -6700 V, providing electric field of 124.7 V/cm. The bias for the cathode was provided using a rack-mounted Bertan Series 205B High Voltage Power Supply.

The field cage is secured from the top. The cathode, which forms the bottom of the field cage, is made with an aluminum honeycomb plate to minimize its weight.  A lightweight cathode minimizes stress on the field cage walls, and reduces the propensity of the cathode to deform. Below the cathode, a custom made voltage step down assembly is used to define and control the electric fields associated with the large voltage difference between the cathode and the outer enclosure.

\begin{figure}
\centering\includegraphics[width=1\linewidth]{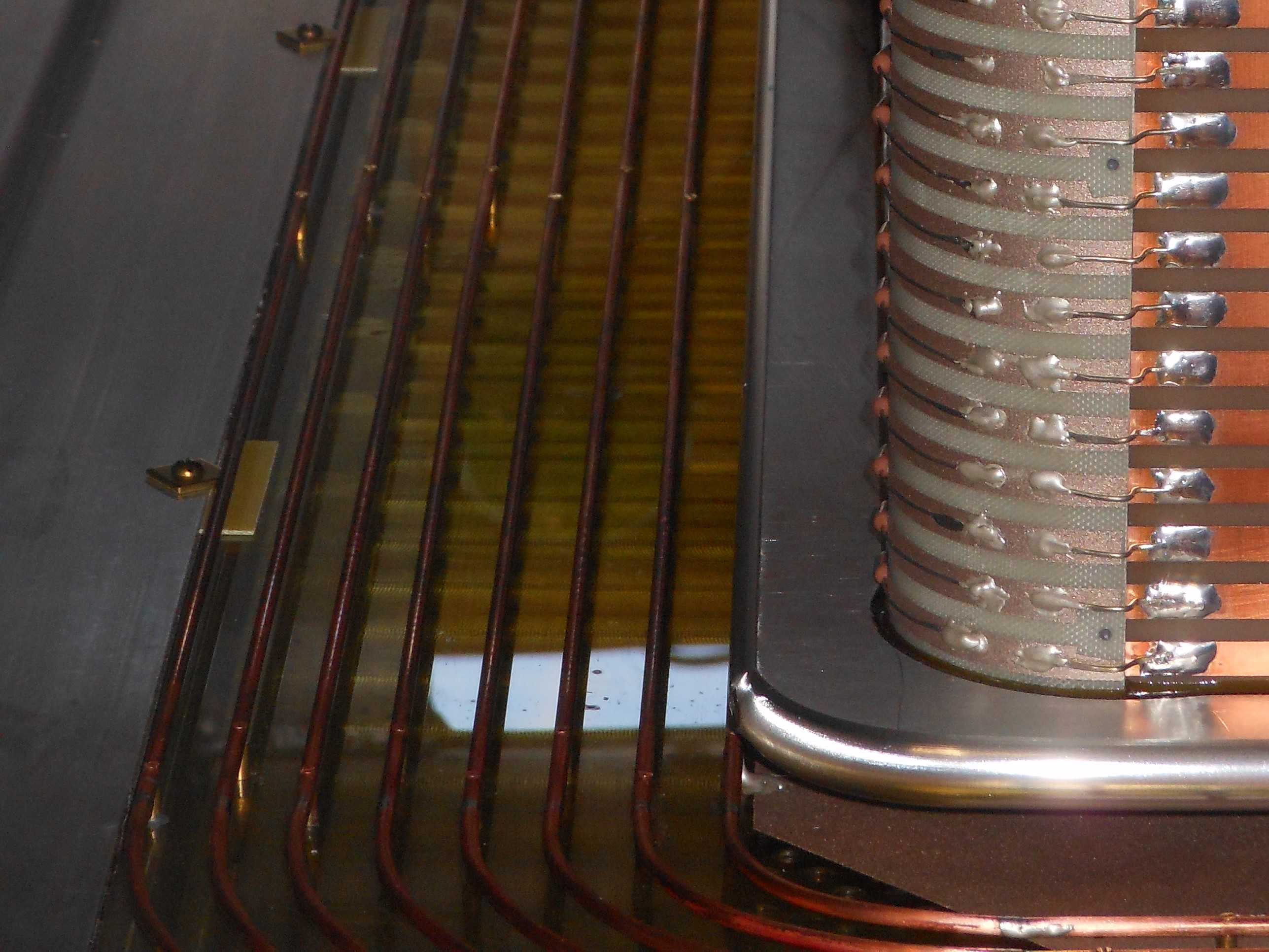}
\caption{Field cage and voltage step down.}
\label{fig:vsd}
\end{figure}

The voltage step-down is built into the bottom plate of the enclosure, and consists of 8 concentric copper loops mounted on a poly-carbonate base. One corner of the voltage step down is shown in Fig.~\ref{fig:vsd}, along with the field cage and cathode.  A conductive surface immediately below the cathode is formed by spray-painting Cho-shield 599 on the poly-carbonate base. This conductive surface is electrically connected to the cathode through spring contacts, and  to the innermost copper ring of the voltage step-down. Each sequentially larger copper ring is connected in series to this innermost ring by 100 M$\Omega$ resistors, and the outermost copper ring is grounded to the enclosure.

During the first experimental campaign, the field cage was filled with P10 gas, which is commonly used for proportional chambers and TPCs such as the EOS TPC and the STAR TPC. The gas was provided to the TPC with a flow rate of 1 L/min. Gas flow was maintained through the field cage via a gas inlet near the upstream end of the cathode and a gas outlet at the top near the downstream end of the pad plane. Gas exiting the field cage was then transmitted by a gas line to the insulation volume and vented from the enclosure through a bubbler filled with 1 cm of mineral oil. This flow system maintained the gas pressure just above atmospheric pressure and allowed the gas volume to be efficiently purged of electronegative gases throughout our measurements. The O$_2$ level was monitored and maintained between 50--70 ppm during the experiment, and the dew-point for the water contaminant in the counter gas was also monitored and controlled so that it remained between -37$^{\circ}$C and -34$^{\circ}$C, corresponding to 170--250 ppm of H$_2$O. The drift velocity of electrons in the gas depends mainly on the ratio of electric field to pressure, $E/p$. Atmospheric pressure was continuously monitored. From this we determined  the typical value of $E/p$ to be 0.167 V/cm/Torr. At this field, P10 is predicted to have a drift velocity of 5.5 cm/\textmu s according to MAGBOLTZ~\cite{biagi2000magboltz} simulation.  During the first campaign of physics measurements, the electron drift velocity was measured to be 5.52$\pm$ 0.02 cm/\textmu s.

\subsection{Target Mechanism}
\begin{figure}
\centering\includegraphics[width=1\linewidth]{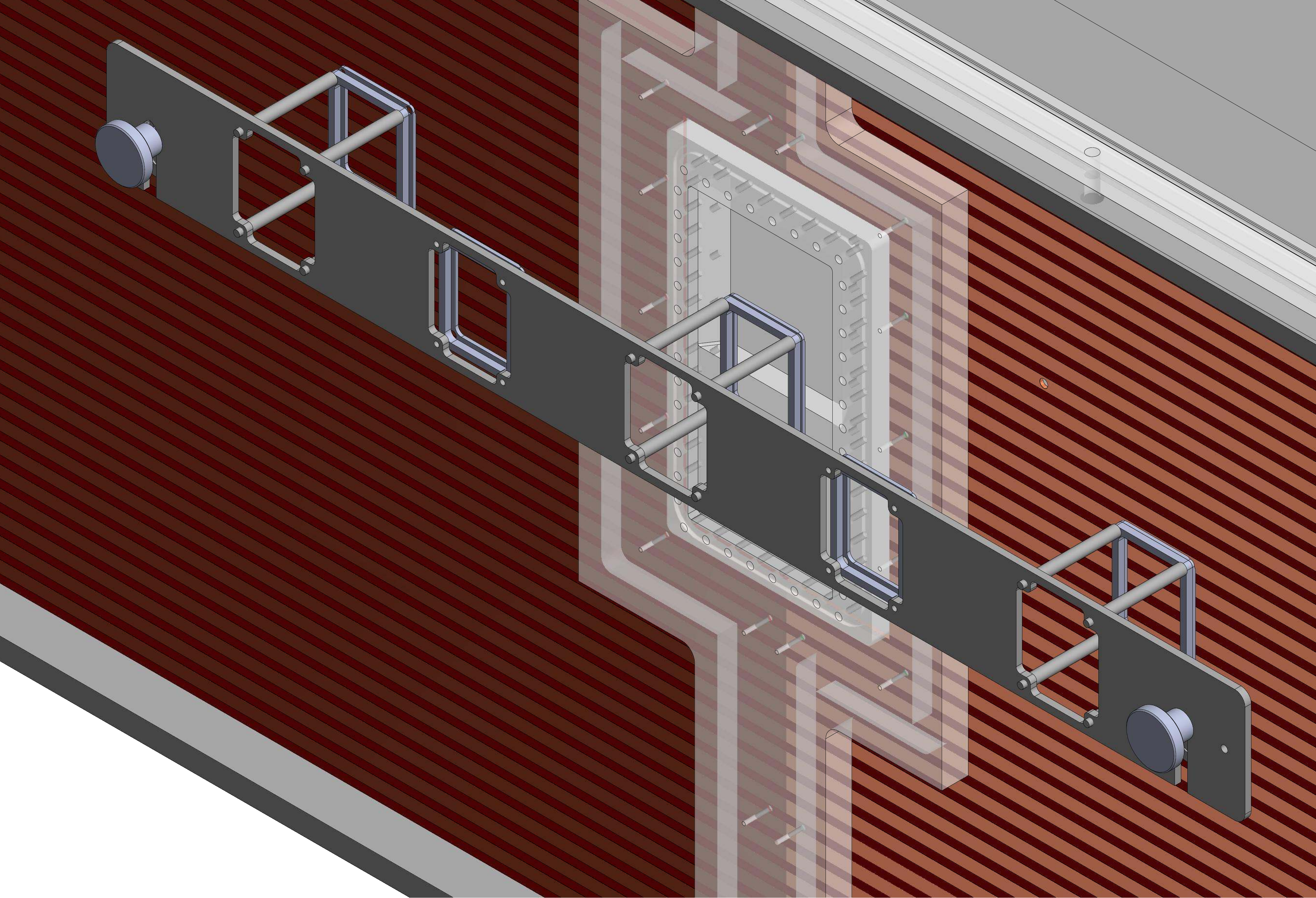}
\caption{Location of target ladder with respect to field cage.}
\label{tgtladder}
\end{figure}

Solid metallic targets form equipotential surfaces that would complicate the transport of electrons from tracks near the target if placed within the drift volume of the TPC.  Since most reaction products of a nucleus-nucleus collision in fixed target mode are emitted at laboratory angles less than $90^{o}$, placing the target just in front of the TPC provides reasonable detection efficiencies for the particles of interest. Therefore, targets are mounted on a target ladder, located outside the field cage, near the entrance window, as shown in Fig.~\ref{tgtladder}. The target ladder has five positions, with three targets mounted on standoffs. These standoffs are necessary to bring the targets as close as possible to the entrance window (which is 3 mm in front of the pad plane origin), thereby maximizing the particle acceptance. The target ladder must be able to move along the $z$-axis to bring the target ladder towards the entrance window, and away from the entrance window prior to moving along the $x$-axis to change targets. These motions are actuated via drive mechanisms from outside the magnetic field and the target positions are calibrated using linear potentiometers. The motion to change target and place it at the correct position is controlled through rotary motion using non-magnetic lead screws. For motion along the $x$-axis, 4 gear pairs are used, and for motion along the $z$-axis, 5 gear pairs are used. The motion is patched through the top plate using commercial rotary O-ring seals for remote control.

\subsection{TPC Enclosure}

\begin{figure}
  \centering
    \includegraphics[width=1\linewidth]{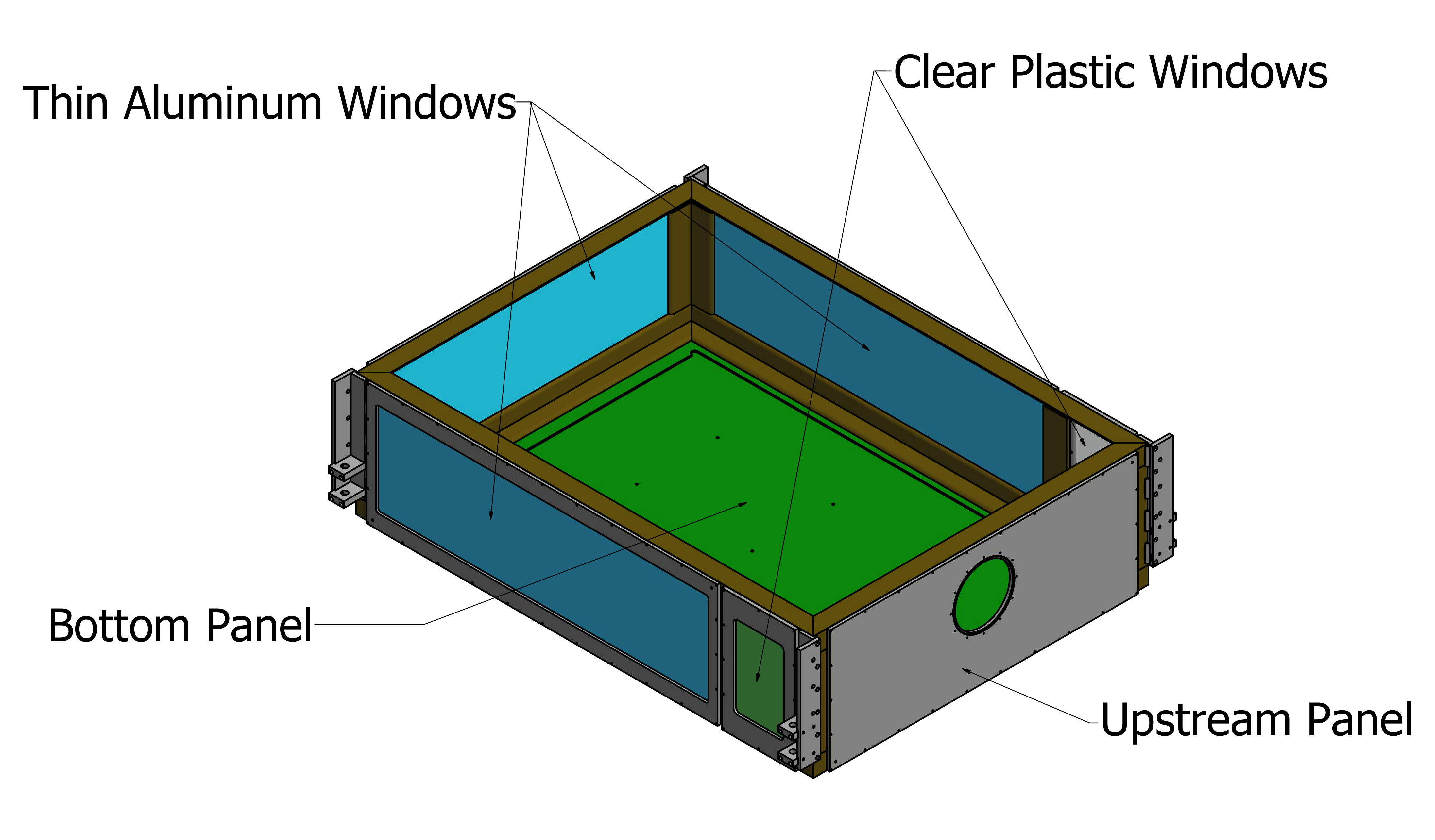}
\caption{TPC enclosure design, without top plate.}
\label{fig:enclosure}
\end{figure}

The TPC provides a gas-tight volume around the field cage, sealed at the top by the rigid top plate. The voltage step down is mounted to the bottom of the enclosure.  The upstream side panel is made of 12.7 mm thick aluminum, with a cutout for an entrance window. An Active Veto~\cite{yan2017veto} shown in Fig.~\ref{TPCexplode} was installed in this cutout, with a 4 \textmu m thick Mylar window. Near the upstream side panels, there are small plastic windows on beam left and right side that provide an unobstructed view of the target motion. Thin aluminum windows (0.95 mm in thickness) cover most of left, right and downstream sides of the enclosure. The thin enclosure and field cage walls allow most of the light charged particles to pass through the walls to the Kyoto~\cite{kaneko2017kyoto} and Katana~\cite{LAS17} scinitillator trigger arrays that are located along the side and downstream walls of the enclosure. The enclosure is shown in Fig.~\ref{fig:enclosure}, with panels and windows labeled. 

\begin{figure}
  \centering
    \includegraphics[width=1\linewidth]{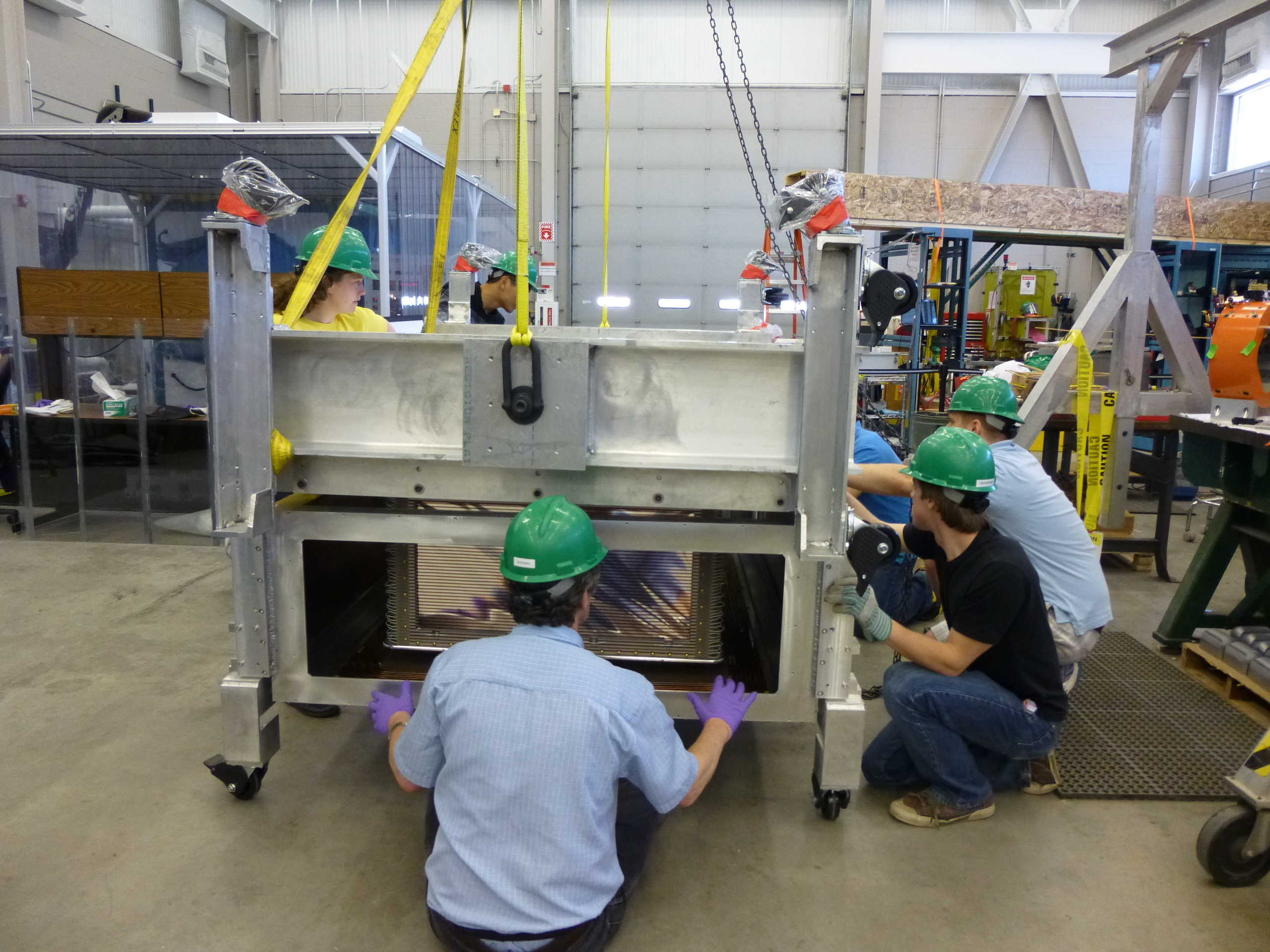}
\caption{Motion chassis attached to enclosure and allowing the TPC to be supported and rotated.}
\label{fig:motion_chassis}
\end{figure}

In addition to the enclosure itself, a versatile motion chassis was constructed to enable the enclosure body and the top plate to be easily and accurately manipulated. The motion chassis served as a support for the top plate when the pad plane and wire planes were constructed. It was designed to allow the top plate to be rotated 90 degrees to pass through narrow doors and to be inverted as shown in Fig.~\ref{fig:motion_chassis} where the top plate and attached field cage are rotated for insertion into the enclosure. When attached to the TPC itself, it allows the TPC to be easily manipulated during assembly, storage and shipping.

\section{S$\pi$RIT TPC performance}
\subsection{Pedestal determination}
\begin{figure}
  \centering
    \includegraphics[width=1\linewidth]{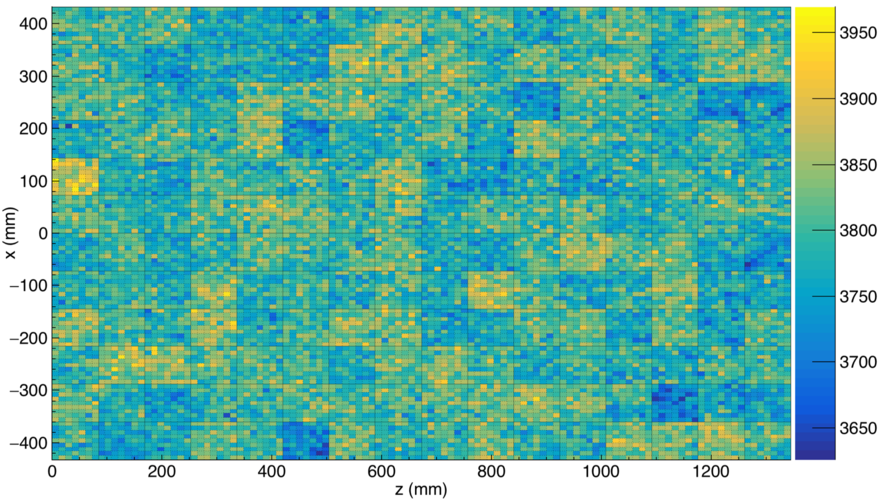}
\caption{The mean pad response over 2000 events with no signals present. This forms the pedestal level for pad signals. }
\label{fig:pedestal}
\end{figure}

The achievable energy loss resolution for any given pad depends on the pedestal and noise. Figure~\ref{fig:pedestal} shows the measured pedestal level for all pads in the pad plane, sampled 2000 times each. The pedestal value is collected by sampling the electronic response when no ionizing radiation is present. The mean pedestal value of all pads was 3796 ADC channels, with mean RMS of 5.3$\pm$0.35 ADC channels. The pedestal stability demonstrates the quality control of the GET fabrication process. After amplification, the signal from the charge induced by the ionization event is negative. The recorded signal from an event is then subtracted from the pedestal to produce a positive pulse, as will be demonstrated in subsequent figures. Thus, the maximum recordable range for each channel is the value of the pedestal, or approximately 3796 ADC channels. During data runs, the first 15 timebuckets (tb) of each pad record the signal prior to the gating grid opening, providing a means to determine the pedestal event-by-event for each pad. From these 15 tb, 10 consecutive tb are selected so as to minimize the RMS value. The mean of the 10 selected timebuckets provides the pedestal value.

\subsection{Cosmic ray Events}
\begin{figure}
  \centering
    \includegraphics[width=1\linewidth]{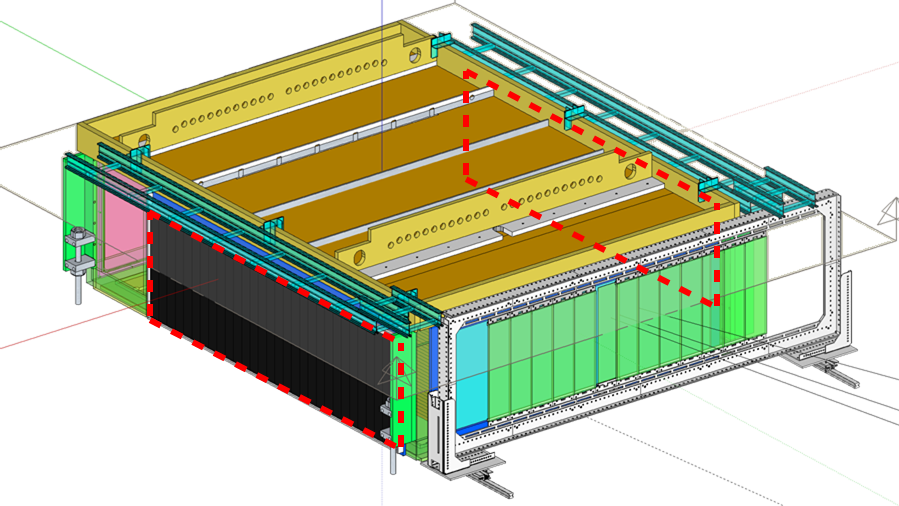}
\caption{TPC with the two-panel Kyoto Multiplicity Array mounted. The panels are shown in black and outlined with red dashed lines. One panel is visible in the figure, and the other panel is on the opposite side of the TPC, with the outline showing its relative location. Modified from~\cite{kaneko2017kyoto}.}
\label{fig:kyoto}
\end{figure}

The S$\pi$RIT TPC detection and tracking abilities were tested with cosmic rays. During this test the TPC was installed in the SAMURAI spectrometer at 0.5 T.  Its data acquisition is triggered by the Kyoto Multiplicity Array~\cite{kaneko2017kyoto} detecting a signal in its left and right array. The Kyoto Multiplicity array consists of 60 plastic scintillator bars, each with dimensions 450$\times$50$\times$10 mm$^3$. These bars are split into two arrays of 30 bars each, which are placed just outside the thin aluminum windows on the left and right sides of the TPC, as shown in Fig.~\ref{fig:kyoto}. Each bar was read out by a Multi Pixel Photon Counter whose output was processed by a VME EASIROC board which provided a multiplicity output to the trigger system~\cite{kaneko2017kyoto}.

\begin{figure}
  \centering
    \includegraphics[width=1\linewidth]{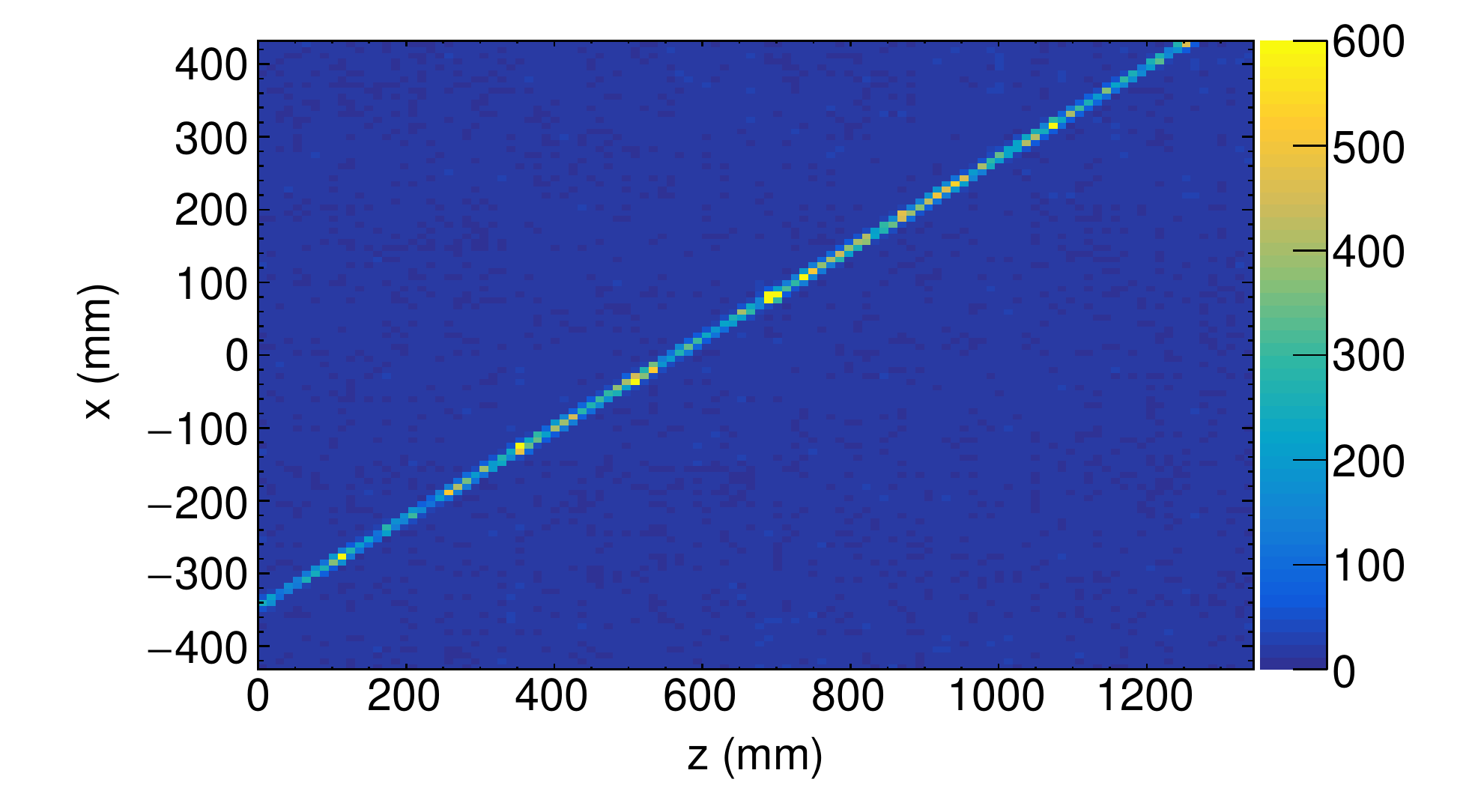}
\caption{A cosmic ray in the S$\pi$RIT TPC, shown from above.}\label{typ_cosmic}
\end{figure}

\begin{figure}
  \centering
    \includegraphics[width=1\linewidth]{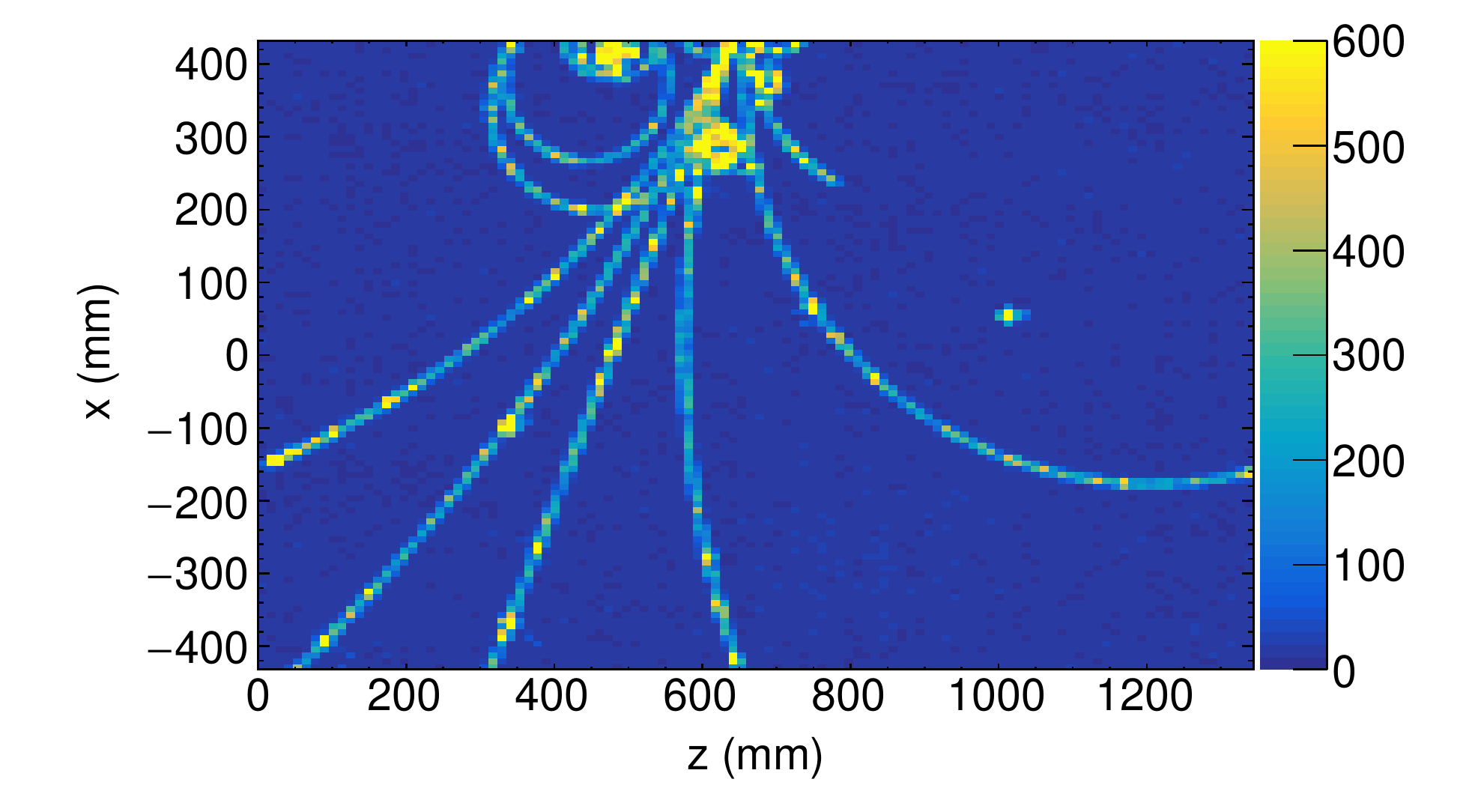}
\caption{A cosmic shower event in the S$\pi$RIT TPC.}\label{typ_shower}
\end{figure}

An example of an event containing one cosmic ray entering the TPC through one half of the Kyoto array and exiting the TPC through the other half of the Kyoto array is shown in Fig.~\ref{typ_cosmic}. In this figure, the pad plane is shown, with each pad forming a pixel. Occasionally, a cosmic array will interact with material in the TPC and produce a shower of particles, as shown in Fig.~\ref{typ_shower}. More images of the cosmic rays and showers can be found in Ref.~\cite{CosmicSite}.

\begin{figure}
  \centering
    \includegraphics[width=1\linewidth]{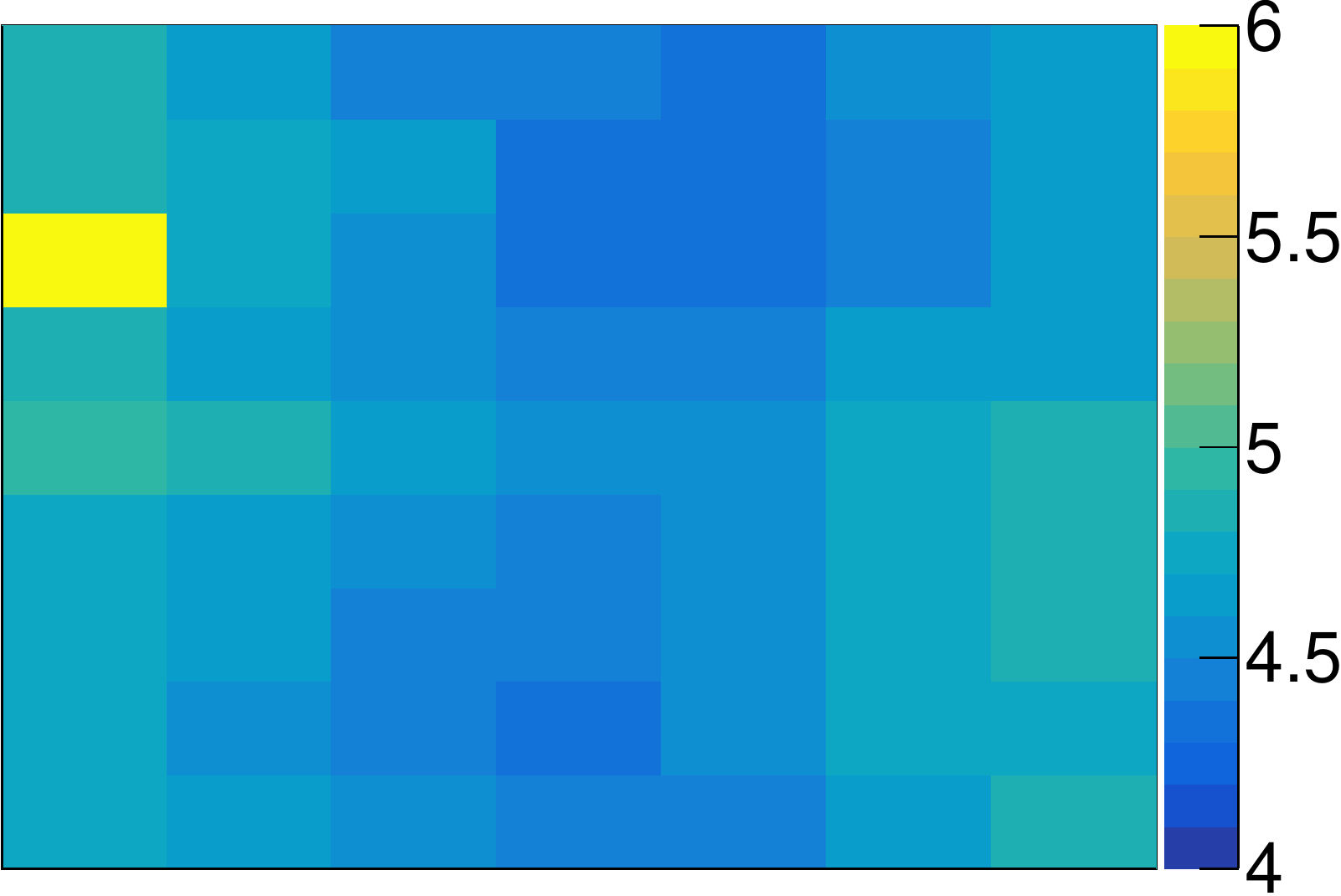}
\caption{Typical noise level of pads within the unit cell, with noise level in ADC channels. The orientation of the pads matches that of Fig.~\ref{ppUnit}.}\label{padnoise}
\end{figure}

The cosmic ray events are ideal to study the response of the TPC to minimum ionizing particles and to perform a measurement of the noise level of the TPC readout, since many pads will have no hits. The average noise level was calculated for each empty pad by taking the maximum deviation from the pedestal value. This is done for 1456 cosmic events collected within a 4 hour period. The noise level clusters in a pattern of unit cells. The average noise level for a unit cell is shown in Fig.~\ref{padnoise}, with the color scale given in Analog-to-Digital-Conversion (ADC) channels, with 1 ADC channel $\approx$ 200 $q_e$. The typical noise level is between 4-5 ADC channels, with a single pad of the unit cell having a noise level between 5.5-6 ADC channels. The noise values are largely consistent with the noise levels expected for the GET system when the input is attached to a capacitance of approximately 25 pf which is typical of our setup. The trajectory of each track is determined by the centroid of a cluster of pads; typically, there are two or more pads in the $x$ and $z$ directions with signals well above the noise for most points on the trajectory, allowing the track centroid to be determined with an accuracy of about 2 mm~\cite{leejw2020}.  

\subsection{Nuclear Reaction Events}
\begin{figure}
  \centering
    \includegraphics[width=1\linewidth]{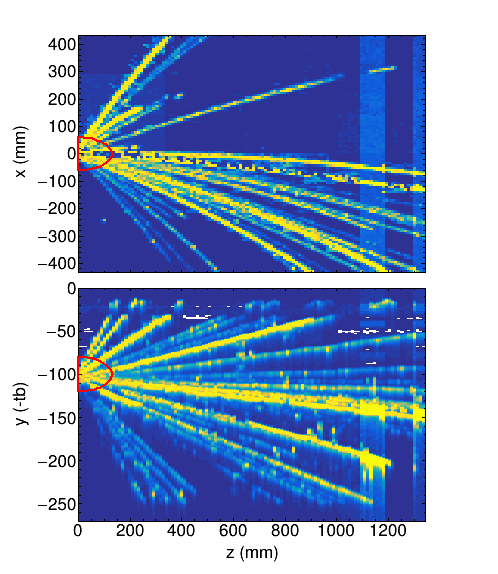}
\caption{A typical event in the TPC, shown from (top) above, and (bottom) the side. The view from above corresponds to the pad plane, with each pixel representing a pad, and the side view showing the time measurement, with each pixel representing a pad length and timebucket width.}\label{typ_event}
\end{figure}

The TPC is used to identify the particles produced in heavy ion collisions. Fig.~\ref{typ_event} shows a typical Sn+Sn collision event in the TPC, viewed from above in the top panel, and viewed from the side in the bottom panel. A number of features can be observed in this event. First, the region near the target, marked with a red half-ellipse, has a high track density, saturating the electronics for many pads in this region and making it impossible to accurately measure the individual tracks in this region.  

Second, the tracks are broken in the vertical direction, as can be seen in the bottom panel between -30 and -20 timebuckets. This is due to the gating grid, which takes 350--400 ns to open and settle down. Above this region, there are portions of these tracks that corresponds to the ionization of particles that physically penetrate through the gating grid and ionize the gas above the gating grid. Third, there are two regions run at reduced gain, between z=1084--1188 mm and z=1296--1344 mm. Finally, there are a number of blue pads in the middle of tracks in regions of high ionization density. These pads are "dead" and provide no useful information in this event. The blue color corresponds to the noise level of the shaper output in the readout electronics. It occurs because the preamps in the readout electronics for these pads are recovering from being saturated by the ionization induced by charged particles that physically penetrated through the gating grid while it was closed. The recovery time for these channels is governed by the preamp fall time of approximately 50 \textmu s.  After 50 \textmu s the preamps recover and these electronic channels are available to take data on the next event. The tracking software is capable of extending  tracks from charged particles past these saturated pads without requiring human intervention.

High amounts of charge on a pad will saturate the charge sensitive amplifier, causing the pad to be insensitive to any further signals for a period of time proportional to the amount of incident charge. The dead pads in Fig.~\ref{typ_event} are most frequently caused by very energetic delta electrons from an un-reacted beam particle that passed through the TPC prior to the recorded event. These electrons spiral upwards along the magnetic field lines through the gating grid while it is closed. Further exploration of this phenomenon is provided in Ref.~\cite{ISOBE201843}. When a pad becomes saturated during an event, information will be recorded up to the point at which it becomes saturated. A saturating signal can be recognized by pulse shape analysis, and this information is incorporated in determining the detection efficiency. The maximum energy loss which can be measured in a single pad is limited by the saturation limit for the charge sensitive amplifier. By incorporating information from nearby pads, however, information on energy loss can be determined even for tracks which saturate the electronics. A software method was developed and implemented in Ref.~\cite{ESTEE2019162509} for the S$\pi$RIT TPC to determine energy loss information for tracks which saturate the electronics. The two regions of lowered gain were used to develop and verify this method, as many of the tracks that saturate the high gain regions do not saturate the low gain region.

The data in Figure~\ref{typ_event} illustrate that positioning the target immediately upstream of the field cage allows most particles to enter the detection volume. Due to this target placement; however, non-interacting beam particles pass directly through the detection volume. These beam particles create slowly drifting  positive ions that form a space charge distribution that distorts the drift field. We have successfully modeled these distortions and applied an inverse-map correction as described in Ref.~\cite{Tsangcf2020}, allowing the proper determination of track trajectories. 

\begin{figure}
  \centering
    \includegraphics[width=1\linewidth]{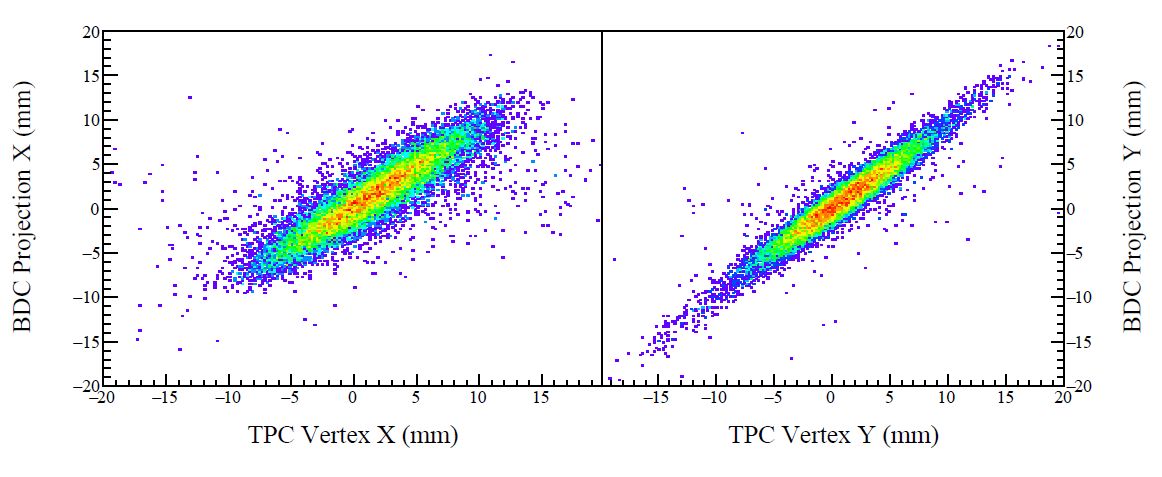}
\caption{Correlations between the event vertex extracted from TPC tracks using RAVE~\cite{Walt2007} and from the intersection of the BDC track and the reaction target. }\label{fig:vertex2}
\end{figure}

The high density region marked by the red half-ellipse is excluded from analysis, as the tracks are too close together to properly separate. Boundary regions directly near any of the field cage edges, including the cathode and gating grid boundaries, were also excluded from analysis. It was observed that tracks in these boundary regions exhibit sharp deflection, which negatively affects event reconstruction. By excluding the boundary and high density regions, we are able to achieve excellent reconstruction of individual tracks, as well as the reconstruction of event characteristics, including the event vertex~\cite{leejw2020}. In the first experimental campaign, the event vertex was determined by two methods. The first involved extrapolating the charged particle tracks in the TPC back to a common vertex. The second involved the use of two two-dimensional Beam Drift Chambers (BDC)~\cite{KOB13} situated in the beam line upstream of the target and using them to project the intersection of the beam with the target. Fig.~\ref{fig:vertex2} shows the correlation between the $x$ and $y$ coordinates of the event vertex obtained from the BDCs and from the TPC using the RAVE vertex determination software~\cite{Walt2007}. The BDC vertex was presently used for track reconstruction because its precision is at the sub-millimeter level. 

\begin{figure}
  \centering
    \includegraphics[width=1\linewidth]{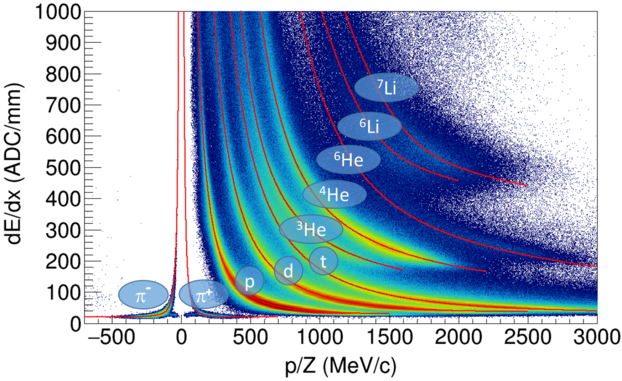}
\caption{PID plot for combined $^{132}$Sn + $^{124}$Sn and $^{108}$Sn + $^{112}$Sn systems. Expected PID lines are drawn up to isotopes of lithium.}\label{tpc_pid}
\end{figure}

The Particle IDentification (PID) is made by plotting the magnetic rigidity $B\rho$ and energy loss dE/dx of each track. Fig.~\ref{tpc_pid} shows the PID from the combined $^{132}$Sn + $^{124}$Sn and the $^{108}$Sn + $^{112}$Sn systems. Both positive and negative pions are clearly visible, as well as hydrogen, helium, and lithium isotopes. Electron and positron lines from $\pi^0$ decays are present, diverging from the pion lines at p/Z$\approx$100 MeV/c, but are not labeled in the figure. The low noise to signal ratio of the electronics allows clear PID line separation. Our saturation corrections, described in Ref.~\cite{ESTEE2019162509} are key to identifying particles with dE/dx > 400 ADC/mm; without corrections, these tracks would be saturated and provide no information~\cite{ESTEE2019162509}. The field corrections for space charge~\cite{Tsangcf2020} are key to accurately measuring momentum.

\section{Conclusions}
The S$\pi$RIT TPC has been constructed, placed within the SAMURAI spectrometer and used successfully in experiments with rare isotope beams at RIBF. A description of key components of the TPC is provided in this work. The S$\pi$RIT TPC performed well, producing good momentum resolution and clear particle identification over a very wide dynamic range in dE/dx for the detected particles. The noise levels were sufficiently low to resolve minimum ionizing particles, with a sufficient gain to resolve Li ions. The particle tracking combined with the corrections for space charge allow precise momentum measurement. Previous publications (References~\cite{ESTEE2019162509,Tsangcf2020} detail the technical solutions developed to correct for space charge effects and achieve a wide dynamic range for energy loss measurement.

Combined, the tracking and energy loss resolution allowed PID separation of the light charged particles between pions and Li ions, which will allow detailed scientific analysis of the beam-target reactions. Solutions to the technical challenges of building the TPC and in using it to measure intermediate energy reactions with rare isotope beams was discussed. The scientific results from experiments using the S$\pi$RIT TPC are the focus of upcoming publications.

\section*{Acknowledgments}
The authors wish to thank the SAMURAI collaboration for their assistance preparing the SAMURAI magnet, as well as helping to install the S$\pi$RIT TPC. We also wish to thank the BigRIPS team for the production and delivery of the rare isotope beams used with the TPC. This work was supported by the U.S. Department of Energy, USA under Grant Nos. DE-SC0014530, DE-NA0002923, DE-FG02-93ER40773, US National Science Foundation, United States Grant No. PHY-1565546, the Japanese MEXT, Japan KAKENHI (Grant-in-Aid for Scientific Research on Innovative Areas) grant No. 24105004, the National Research Foundation of Korea under grant Nos. 2016K1A3A7A09005578, 2018R1A5A1025563, the Polish National Science Center (NCN), Poland, under contract Nos. UMO-2013/09/B/ST2/04064, UMO-2013/-10/M/ST2/00624, and the Robert A. Welch Foundation, United States (A-1266). The computing resources for analyzing the data was supported by the HOKUSAI-GreatWave system at RIKEN, the Institute for Cyber-Enabled Research (ICER) cluster at Michigan State University, and the EMBER cluster at the NSCL.

\bibliographystyle{elsarticle-num}
\bibliography{main}

\end{document}